%% file: smartwatts-paper.tex
\documentclass[conference]{IEEEtran}
\IEEEoverridecommandlockouts
\usepackage{cite}
\usepackage{amsmath,amssymb,amsfonts}
\usepackage{algorithmic}
\usepackage{graphicx}
\usepackage{textcomp}
\usepackage{paralist}
\usepackage{xcolor}

\usepackage[normalem]{ulem}
\usepackage{hyperref}
\usepackage{booktabs}

\pdfoutput=1

\newcommand{\SYS}[0]{\textsc{Smart\-Watts}}
\newcommand{\SYSERROR}[0]{3.5\,\%}
\newcommand{\SYSURL}[0]{\url{https://github.com/powerapi-ng/smartwatts-formula}}

\begin{document}

\author{
    \IEEEauthorblockN{\emph{Guillaume} \textbf{Fieni}}
    \IEEEauthorblockA{Univ. Lille / Inria\\
    France\\
    \small \tt guillaume.fieni@univ-lille.fr}
\and
    \IEEEauthorblockN{\emph{Romain} \textbf{Rouvoy}}
    \IEEEauthorblockA{Univ. Lille / Inria / IUF\\
    France\\
    \small \tt romain.rouvoy@univ-lille.fr}
\and
    \IEEEauthorblockN{\emph{Lionel} \textbf{Seinturier}}
    \IEEEauthorblockA{Univ. Lille / Inria\\
    France\\
    \small \tt lionel.seinturier@univ-lille.fr}
}

\title{\SYS{}: Self-Calibrating Software-Defined~Power~Meter for Containers}

\maketitle

\thispagestyle{plain}
\pagestyle{plain}

\input{sections/abstract}

\input{sections/introduction}
\input{sections/related-works}
\input{sections/contribution}

\input{sections/implementation}

\input{sections/validation}
\input{sections/conclusion}
\section*{Acknowledgement}
The authors would like to thank Jo\"el Penhoat for his insightful feedbacks on this version of the paper.

\bibliographystyle{IEEEtran}
\bibliography{IEEEabrv,references}

\end{document}

%% file: sections/abstract.tex
\begin{abstract}
Fine-grained power monitoring of software activities becomes unavoidable to maximize the power usage efficiency of data centers.
In particular, achieving an optimal scheduling of containers requires the deployment of software-defined power~meters to go beyond the granularity of hardware power monitoring sensors, such as \emph{Power Distribution Units} (PDU) or Intel's \emph{Running Average Power Limit} (RAPL), to deliver power estimations of activities at the granularity of software~containers.
However, the definition of the underlying power models that estimate the power consumption remains a long and fragile process that is tightly coupled to the host machine.

To overcome these limitations, this paper introduces \SYS{}: a lightweight power monitoring system that adopts online calibration to automatically adjust the CPU and DRAM power models in order to maximize the accuracy of runtime power estimations of containers.
Unlike state-of-the-art techniques, \SYS{} does not require any \emph{a~priori} training phase or hardware equipment to configure the power models and can therefore be deployed on a wide range of machines including the latest power optimizations, at no cost.
\end{abstract}

\begin{IEEEkeywords}
    Energy, Containers, Power model
\end{IEEEkeywords}

%% file: sections/introduction.tex
\section{Introduction}
Modern data centers are continuously trying to maximize the \emph{power usage efficiency} (PUE) of their hardware and software infrastructures to reduce their operating cost and eventually their carbon emission.
While physical power~meters offer a suitable solution to monitor the power consumption of physical servers, they fail to support the energy profiling at a finer granularity: dealing with the software services that are distributed across such infrastructures.
To overcome this limitation, software-defined power~meters build on power models to estimate the power consumption of software artifacts in order to identify potential energy hotspots and leaks in software systems~\cite{Noureddine:2015} or improve the management of resources~\cite{DBLP:conf/eurosys/SnowdonSPH09}.
However, existing software-defined power~meters are integrating power models that are statically designed, or learned prior to any deployment in production~\cite{Colmant:2018,Lebeane:2015}.
This may result in inaccuracies in power estimations when facing unforeseen environments or workloads, thus affecting the exploitation process.
As many distributed infrastructures, such as clusters or data~centers, have to deal with the scheduling of unforeseen jobs, in particular when handling black-box virtual machines, we can conclude that the adoption of such static power models~\cite{Colmant:2015} has to be considered as inadequate in production.
We therefore believe that the state-of-the-art in this domain should move towards the integration of more dynamic power models that can adjust themselves at runtime to better reflect the variation of the underlying workloads and to cope with the potential heterogeneity of the host machines.

In this paper, we therefore introduce \SYS{}, as a self-calibrating software-defined power~meter that can automatically adjust its CPU and DRAM power models to meet the power accuracy requirements of monitored software containers.
Our approach builds on the principles of sequential learning principles and proposes to exploit coarse-grained power monitors like \emph{Running Power Average Limit} (RAPL), which is commonly available on modern Intel's and AMD's micro-architecture generations, to control the estimation error.
We have implemented \SYS{} as an open~source power~meter to integrate our self-calibrating approach, which is in charge of automatically adjusting the power model whenever some deviation from the ground truth is detected.
When triggered, the computation of a new power model aggregates the past performance metrics from all the deployed containers to infer a more accurate power model and to seamlessly update the software-defined power~meter configuration, without any interruption.
The deployment of \SYS{} in various environments, ranging from private clouds to distributed HPC clusters, demonstrates that \SYS{} can ensure accurate real-time power estimations (less than \SYSERROR{} of error on average, at a frequency of 2\,Hz) at the granularity of processes, containers and virtual machines.
Interestingly, the introduction of sequential learning in software-defined power~meters eliminates the learning phase, which usually last from minutes to hours or days, depending on the complexity of the hosting infrastructure~\cite{Colmant:2015,Colmant:2018}.

Additionally, our software-defined approach does not require any specific hardware investment as \SYS{} can build upon embedded power sensors, like RAPL, whenever they are available.
The code of \SYS{} is made available online as open-source software\footnote{\SYSURL{}} to encourage its deployment at scale and to leverage the adoption and reproduction of our results.
The key contributions of this paper can therefore be summarized as follows:
\begin{compactenum}
  \item a self-calibrating power modelling approach,
  \item CPU \& DRAM models supporting power states,
  \item an open source implementation of our approach,
  \item an assessment on container-based environments.
\end{compactenum}

In the remainder of this paper, we start by providing some background on state-of-the-art power models and their limitations (cf. Section~\ref{sec:relatedwork}) prior to introducing our contribution (cf. Section~\ref{sec:contribution}).
Then, we detail the implementation of \SYS{} as an extension of the \textsc{BitWatts} middleware framework (cf. Section~\ref{sec:implementation}) and we assess its validity on three scenarios (cf. Section~\ref{sec:validation}).
We conclude and provide some perspectives for this work in Section~\ref{sec:conclusion}.

%% file: sections/related-works.tex
\section{Related Work}\label{sec:relatedwork}

\subsection{Hardware Power~Meters}\label{sec:meters}
Over the years, hardware power~meters have evolved to deliver hardware-level power measurements with different levels of granularity, from physical machines to electronic components.

\textsc{WattProf}~\cite{Rashti:2015} power monitoring platform supports the profiling of \emph{High Performance Computing} (HPC) applications.
This solution is based on a custom board, which can collect raw power measurements from various hardware components (CPU, disk, memory, etc.) from sensors connected to power lines.
The board can connect up to $128$ sensors that can be sampled at up to $12\,KHz$.
As in~\cite{Ge:2010}, the authors argue that this solution is able to perform per-process power estimation, but they only validate their approach while running a single application.

\textsc{WattWatcher}~\cite{Lebeane:2015} is a tool that can characterize workload energy power consumption.
The authors use several calibration phases to build a power model that fits a CPU architecture.
This power model uses a predefined set of \emph{Hardware Performance Counters} \textsc{(HwPC)} as input parameters.
As the authors use a special power model generator that can target any CPU architecture, which has be to carefully described.

\textsc{RAPL}~\cite{Rotem:2012} offers specific \emph{hardware performance counters} (\textsc{HwPC}) to report on the energy consumption of the CPU since the ``Sandy Bridge`` micro-architecture for Intel (2011) and ``Zen`` for AMD (2017).
Intel divides the system into domains (\textsf{PP0, PP1, PKG, DRAM}) that report the energy consumption according to the requested context.
The \textsf{PP0} domain represents the core activity of the processor (cores\,+\,L1\,+\,L2\,+\,L3), the \textsf{PP1} domain the uncore activities (LLC, integrated graphic cards, etc.), and \textsf{PKG} represents the sum of \textsf{PP0} and \textsf{PP1}, and the \textsf{DRAM} domain exhibits the DRAM energy consumption.
Desrochers~\emph{et al.} demonstrate the accuracy of the DRAM power estimations of RAPL, especially on Intel Xeon processors~\cite{Desrochers:2016}.

\subsection{Software-Defined Power~Meters}\label{sec:models}
To get rid of the hardware cost imposed by the above solutions, the design of power models has been regularly considered by the research community over the last decade, in particular for CPU~\cite{Bellosa:2000,Colmant:2015,Kansal:2010,McCullough:2011,Versick:2013}.
Notably, as most architectures do not provide fine-grained power measurement capabilities, McCullough~\emph{et~al.}~\cite{McCullough:2011} argue that power models are the first step towards enabling dynamic power management for power proportionality at all levels of a system.

While standard operating system metrics (CPU, memory, disk, or network), directly computed by the kernel, tend to exhibit a large error rate due to their lack of precision~\cite{Kansal:2010,Versick:2013}, \textsc{HwPC} can be directly gathered from the processor (\emph{e.g.}, number of retired instructions, cache misses, non-halted cycles).
Modern processors provide a variable number of \textsc{HwPC} events, depending on the generation of the micro-architectures and the model of the CPU. 
As shown by Bellosa~\cite{Bellosa:2000} and Bircher~\cite{Bircher:2007}, some \textsc{HwPC} events are highly correlated with the processor power consumption, while the authors in~\cite{Rivoire:2008} concluded that not all HPC are relevant, as they may not be directly correlated with dynamic power.

Power modeling often builds on these raw metrics to apply learning techniques~\cite{Bertran:2010} to correlate the metrics with hardware power measurements using various regression models, which are so far mostly linear~\cite{McCullough:2011}.
Three key components are commonly considered to train a power model:
\begin{inparaenum}[\em a)]
    \item the workload(s) to run during sampling,
    \item the minimal set of input parameters, and 
    \item the class of regression to use~\cite{Bertran:2010, Zhai:2014, Zamani:2012, Dolz:2015}.
\end{inparaenum}

The workloads used along the training phase have to be carefully selected to capture the targeted system.
In this domain, many benchmarks have been considered, but they are mostly 
\begin{inparaenum}[\em a)]
    \item designed for a given architecture~\cite{Bertran:2010,Isci:2003}, 
    \item manually selected~\cite{Bircher:2005,Colmant:2015,Contreras:2005,Dolz:2015,Li:2003,Yang:2014,Zamani:2012,Zhai:2014}, or even 
    \item private~\cite{Zhai:2014}.
\end{inparaenum}
Unfortunately, this often leads to the design of power models that are tailored to a given processor architecture and manually tuned (for a limited set of power-aware features)~\cite{Bertran:2010,Bircher:2005,Isci:2003,Li:2003,Lim:2010,Yang:2014,Zhai:2014,Shen:2013}.

\subsection{Limitations \& Opportunities}
To the best of our knowledge, the state of the art in hardware power~meters often imposes hardware investments to provide power measurements with an high accuracy, but a coarse granularity, while software-defined power~meters target fine-grained power monitoring, but often fail to reach high accuracy on any architecture and/or workload.

This paper clearly differs from the state of the art by providing an open source, modular, and self-adaptive implementation of a self-calibrating software-defined power~meter: \SYS{}.
As far as we know, our implementation is the first to deliver both CPU and DRAM power estimations at runtime for any software packaged as processes, containers or virtual machines.
Unlike existing approaches published in the literature, the approach we describe is \emph{i)}~architecture agnostic, \emph{ii)}~processor aware, and \emph{iii)}~dynamic.
So far, the state of the art fails to deploy software-defined power~meters in productions because \emph{i)} the model learning phase can last from minutes to days, \emph{ii)} the power models are often bound to a specific context of execution that do not take into account hardware energy-optimization states, and \emph{iii)} the reference power measurement requires specific hardware to be installed on a large amount of nodes.
This therefore calls for methods that can automatically adapt to the hardware and workload diversities of heterogenous environments in order to maintain the accuracy of power measurements at scale.

%% file: sections/contribution.tex
\section{\SYS{} Power Monitoring}\label{sec:contribution}
We therefore propose to support self-calibrating power models that leverage \emph{Reference Measurements} and \emph{Hardware Performance Counters} (\textsc{HwPC}) to estimate the power consumption at the granularity of software containers along multiple resources: CPU and DRAM.
More specifically, our contribution builds upon two widely available system interfaces: \emph{RAPL} to collect baseline measurements for CPU and DRAM power consumptions, as well as Linux's \emph{perf\_events} interface to capture the \emph{Hardware Performance Counters} (\textsc{HwPC}) events used to estimate the per-container power consumption from resource-specific power models, which are adjusted at runtime.

\subsection{Overview of \SYS{}}
\begin{figure}
  \includegraphics[width=\linewidth]{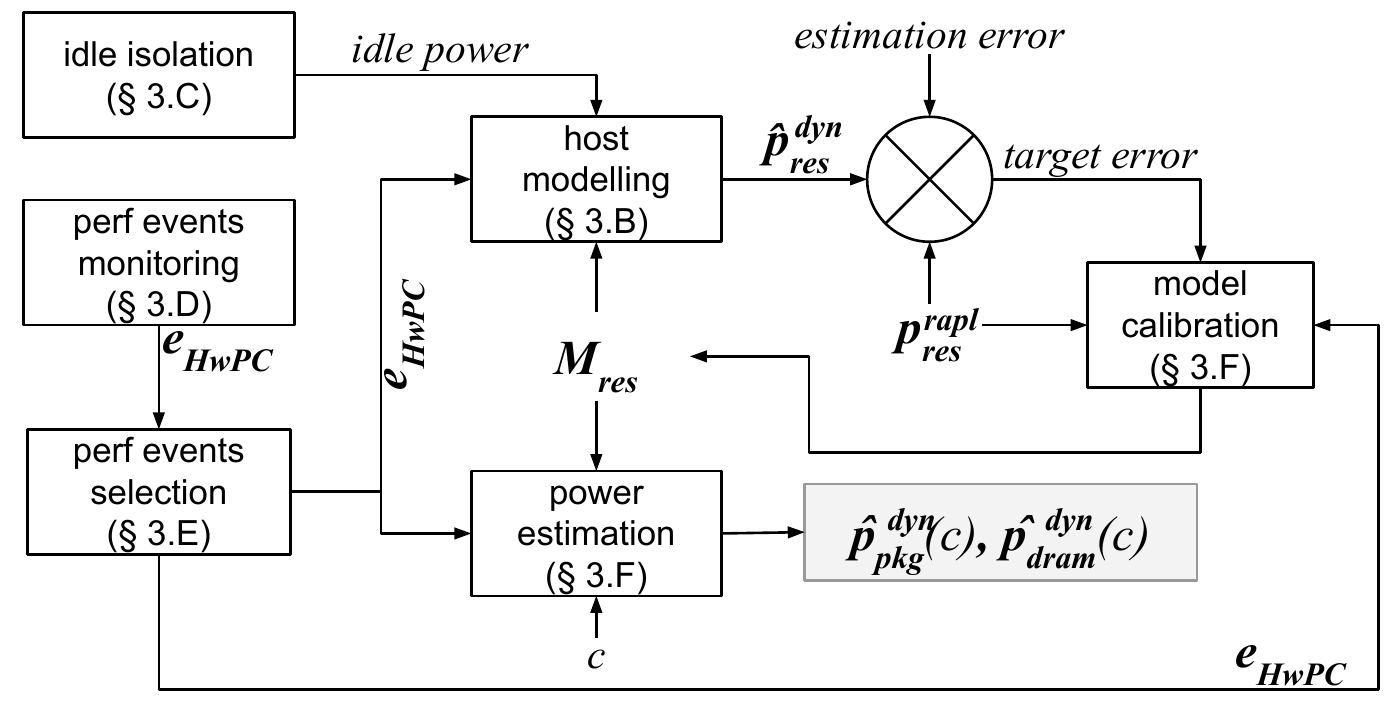}
  \caption{Overview of \SYS{}}
  \label{fig:overview}
\end{figure}

Figure~\ref{fig:overview} introduces the general architecture of \SYS{}.
\SYS{} manages at runtime a set of self-calibrated power models ($M^f_{res}$) for each \emph{power-monitorable} resource $res$ (\emph{e.g.}, CPU, DRAM).
These power models are then used by \SYS{} to estimate the power consumptions of \emph{i)} the host $\hat{p}_{res}$ and \emph{ii)} all the hosted containers $c$: $\hat{p}_{res}(c)$.

\SYS{} uses $\hat{p}_{res}$ to continuously assess the accuracy of the managed power models ($M^f_{res}$) and to ensure that the estimated power consumption does not diverge from the baseline measurements reported by RAPL ($p^{rapl}_{res}$, cf. Section~\ref{sec:model}).
Whenever the estimated power consumption error ($\epsilon_{res}$) diverges from the baseline measurements beyond a configured threshold, \SYS{} automatically triggers a new online calibration process of the diverging power model to better match the current input workload.%

To better capture the dynamic power consumption of the host, \SYS{} needs to isolate its static consumption.
To do so, we use a dedicated component that activates when the machine is at rest---\emph{e.g.}, after booting (cf. Section~\ref{sec:idle})---to monitor the power activity of the host.

In addition to the static constant, \SYS{} estimates the power consumption of the host from a set of raw input values that refers to \textsc{HwPC} events, which are selected \emph{at runtime} (cf. Section~\ref{sec:hpc}).

This design ensures that \SYS{} keeps adjusting its power models to maximize the accuracy of power estimations.
Therefore, unlike the state-of-the-art power monitoring solutions, \SYS{} does not suffers from estimation errors due to the adoption of an inappropriate power model as it autonomously optimizes the underlying power model whenever a accuracy anomaly is detected.

\subsection{Modelling the Host Power Consumption}\label{sec:model}
For each resource $res \in \{pkg, dram\}$ exposed by the RAPL interface, the associated power consumption $p^{rapl}_{res}$ can be modelled as:
\begin{equation}
p^{rapl}_{res} = p^{static}_{res} + p^{dyn}_{res}
\end{equation}
where $p^{static}_{res}$ refers to the static power consumption of the monitored resource (cf. Section~\ref{sec:idle}), and $p^{dyn}_{res}$ reflects the dynamic power dissipated by the processor along the sampling period.

Then, we can compute a power model $M^f_{res} = [\alpha_0, \cdots, \alpha_n]$ that correlates, for a given frequency $f$ (among available frequencies $F$, cf. Section~\ref{sec:monitoring}), the dynamic power consumption ($\hat{p}^{dyn}_{res}$) to the raw metrics reported by a set of of \emph{Hardware Performance Counter} (HwPC) events (cf. Section~\ref{sec:hpc}), $E^{f}_{res} = [e_0, \dots , e_n]$: 
\begin{equation}
\exists f \in F,~\hat{p}^{dyn}_{res} = M^f_{res} \boldsymbol{\cdot} E^f_{res}
\end{equation}

We build $M^f_{res}$ from a Ridge regression---a linear least squares regression with l2 regularization---applied on the past $k$ samples $S^f_k = \langle p^{dyn}_{res}, E^f_{res} \rangle$, with $p^{dyn}_{res} = p^{rapl}_{res} - p^{static}_{res}$.
By comparing $p^{dyn}_{res} + p^{static}_{res}$ with $p^{rapl}_{res}$, we can continuously estimate the error $\varepsilon_{res} = \mid p^{dyn}_{res} - \hat{p}^{dyn}_{res} \mid$ from estimated values in order to monitor the accuracy of the power model $M^f_{res}$.
Whenever the error exceeds a given threshold set by the administrator, a new power model is generated for the frequency $f$ by integrating the latest samples.

\subsection{Isolating the Static Power Consumption}\label{sec:idle}
Isolating the static power consumption of a node is a challenging issue as it requires to reach a quiescient state in order to capture the power consumption of the host at rest.
To capture this information, we designed and implemented a power logger component that runs as a lightweight daemon with low priority that periodically logs the package and DRAM power consumptions reported by RAPL.
Then, we compute the \emph{median} value and the \emph{interquartile range} (IQR) from gathered measurements to define the $p^{static}_{res}$ constant as : $p^{static}_{res} = median_{res} - 1.5 \times IQR_{res}$.
This approach intends to filter out outliers reported by RAPL, including periodic measurement errors we observed, and to consider the lowest power consumption observed along a given period of time.

By default, \SYS{} assumes that the static consumption of the host does not requires to be spread across the active containers.
However, other power accounting policies can be implemented.
For example, by reporting an empty static consumption, \SYS{} will share it across the running containers depending on their activity.

\subsection{Monitoring Power States \& \textsc{HwPC} Events}\label{sec:monitoring}
As previously introduced, the accuracy of a power model $M^f_{res}$ strongly depends on \emph{i)} the selection of relevant input features (\textsc{HwPC} events $e_n$) and \emph{ii)} the acquisition of input values that are evenly distributed along the reference power consumption range.
This is one of the reasons why the input workloads used in standard calibration phases are often critical to capture an accurate power model that reflects the power consumption of a host for a given class of applications.
\SYS{} rather promotes a self-calibrating approach that does not impose the choice of a specific benchmark or workload, but exploits the ongoing activity variations of the host machine to continuously adjust its power models.
To achieve this, \SYS{} monitors selected sets of \textsc{HwPC} events and stores the associated samples in memory.
To better deal with the power features of hardware components, we group the input samples per operating frequency.
This allows to calibrate frequency-specific power models when an estimation arises, with the goal to converge automatically to a stable and precise power model over the time.

By balancing the samples along the range of frequencies operated by the processor, \SYS{} ensures that the power model learning phase does not overfit the current context of execution, which may lead to the generation of unstable power models, thus impacting the accuracy of the power measurements.
The sampling tuples $S^f_k$ are grouped into memory as frequency layers $L^{f}_{res}=[S^f_0, ..., S^f_n]$, which are the raw features we maintain to build $M^f_{res}$.

To store the samples in the layer corresponding to the current frequency of the processor, \SYS{} compute the average running frequency as follows:
\begin{equation}
  F_{avg} = F_{base} * \frac{\Delta~{\tt APERF}}{\Delta~{\tt MPERF}}
\end{equation}
where $F_{base}$ is the processor base frequency constant extracted from the \emph{Model Specific Registers} (MSR) \texttt{PLATFORM\_INFO}.
\texttt{APERF} and \texttt{MPERF} are MSR-based counters that increment at the current and maximum frequencies, respectively.
These counters are continuously updated, hence they report on a precise average frequency without consuming the limited \textsc{HwPC} slots.
Interestingly, the performance power states, such as P-states and Turbo~Boost, will be accounted by these counters as they act mainly on the frequency of the core in order to boost the performance.
The idle optimization states (C-states) will also be accounted as they mainly reduce of the average frequency of the core towards its \emph{Max Efficiency Frequency} before being powered-down.

\subsection{Selecting the Correlated \textsc{HwPC} Events}\label{sec:hpc}
The second challenge of \SYS{} consists in selecting at runtime the relevant \textsc{HwPC} events that can be exploited to accurately estimate the power consumption.
To do so, we list the available events exposed by the host's \emph{Performance Monitoring Units} (PMU) and we evaluate their correlation with the power consumption reported by RAPL.
Instead of testing each available \textsc{HwPC} events, we narrow the search using the PMU associated to the modelled component---\emph{i.e.}, we consider the \textsc{HwPC} events from the \textsf{core} PMU to model the PKG power consumption.
As reference events, we consider \texttt{unhalted-cycles} for the package and \texttt{llc-misses} for the DRAM, which are the standard \textsc{HwPC} events available across many processor architectures, and have been widely used by the state of the art to design power models~\cite{Colmant:2015,Colmant:2018,Lebeane:2015}.
To elect a \textsc{HwPC} event as a candidate for the power model, we first compute the Pearson coefficient $r_{e,p}$ for $n$ values reported by each monitored \textsc{HwPC} event $e$ and the base power consumption $p$ reported by RAPL:
\begin{equation}\label{eq:pearson}
r_{e,p} = \frac{\sum\limits_{i=1}^n{(e_i - \overline{e})~(p_i - \overline{p})}}{\sqrt{\sum\limits_{i=1}^n{(e_i - \overline{e})^2}}~\sqrt{\sum\limits_{i=1}^n{(p_i - \overline{p})^2}}}
\end{equation}

Then, \SYS{} stores the list of \textsc{HwPC} events that exhibit a better correlation coefficient $r$ than the baseline event for DRAM and PKG.
This list of elected \textsc{HwPC} events is further used as input features to implement the PKG and DRAM power models exploited by \SYS{}.

\subsection{Estimating the Container Power Consumption}
Given that we learn the power model $M^f_{res}$ from aggregated events, $E^f_{res} = \sum_{c \in C} E^f_{res}(c)$, we can predict the power consumption of any container $c$ by applying the inferred power model $M^f_{res}$ at the scale of the container's events $E^f_{res}(c)$:
\begin{equation}
\exists f \in F,~\forall c \in C,~\hat{p}^{dyn}_{res}(c) = M^f_{res} \boldsymbol{\cdot} E^f_{res}(c)
\end{equation}

In theory, one can expect that $\hat{p}^{dyn}_{res}\overset{!}{=}p^{dyn}_{res}$ if the model perfectly estimates the dynamic power consumption but, in practice, the predicted value may introduce an error $\varepsilon_{res} = \mid p^{dyn}_{res} - \hat{p}^{dyn}_{res} \mid$.
Therefore, we cap the power consumption of any container $c$ as:
\begin{equation}
\forall c \in C,~\lceil\hat{p}^{dyn}_{res}(c)\rceil = \frac{p^{dyn}_{res} \times \hat{p}^{dyn}_{res}(c)}{\hat{p}^{dyn}_{res}}
\end{equation}
to ensure that $p^{dyn}_{res} = \sum_{c \in C}\lceil\hat{p}^{dyn}_{res}(c)\rceil$, thus avoiding potential outliers.
Thanks to this approach, we can also report on the confidence interval of the power consumption of containers by scaling down the observed global error:
\begin{equation}
\forall c \in C,~\varepsilon_{res}(c) = \frac{\hat{p}^{dyn}_{res}(c)}{\hat{p}^{dyn}_{res}} \times \varepsilon_{res}
\end{equation}

In the following sections, we derive and implement the above formula to report on the power consumption of $pkg$ and $dram$ resources.
Our empirical evaluations report on the capped power consumptions for $pkg$ ($\lceil\hat{p}^{dyn}_{pkg}\rceil$) and $dram$ ($\lceil\hat{p}^{dyn}_{dram}\rceil$), as well as the associated errors $\varepsilon_{pkg}$ and $\varepsilon_{dram}$, respectively.

%% file: sections/implementation.tex
\section{Implementation of \SYS{}}\label{sec:implementation} %
\begin{figure}
    \includegraphics[width=\linewidth]{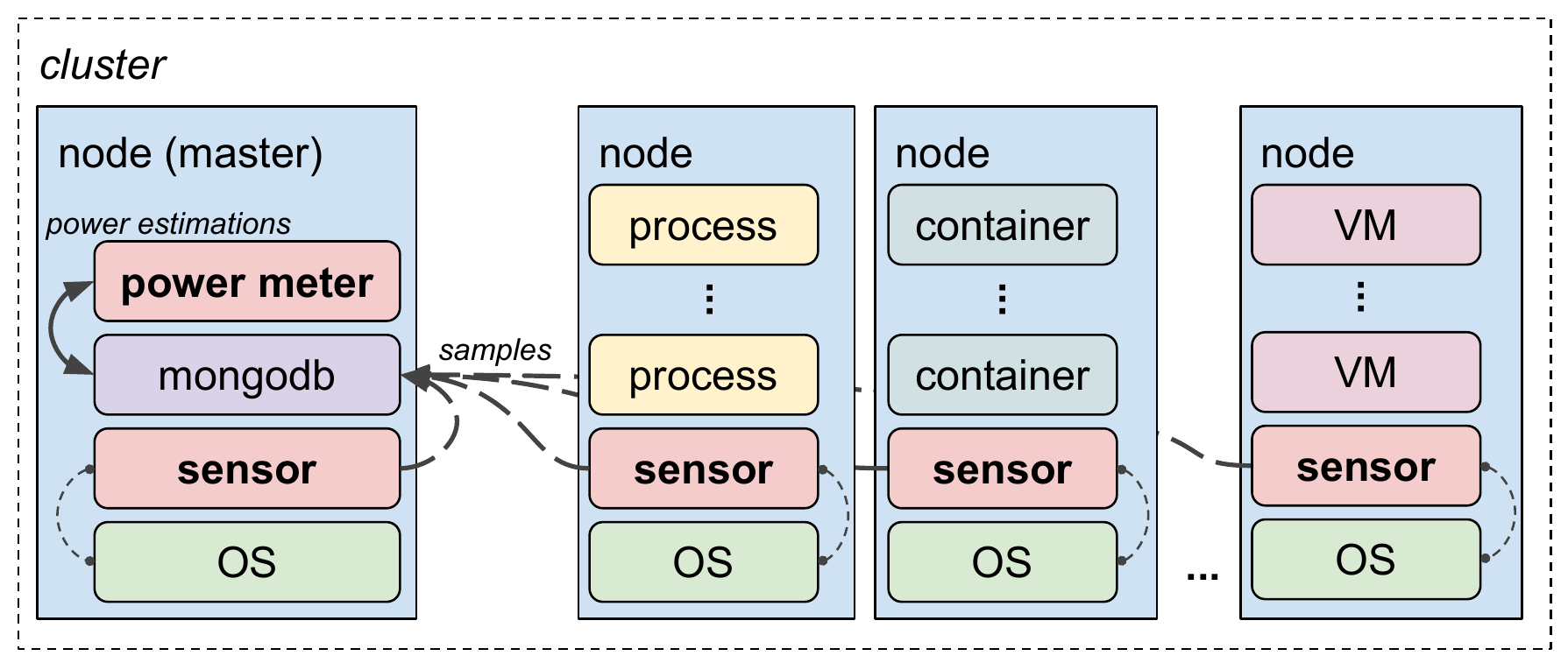}
    \caption{Deployment of \SYS{}}
    \label{fig:deployment}
\end{figure}

We implemented \SYS{} as a modular software system that can run atop a wide diversity of production environments.
As depicted in Figure~\ref{fig:deployment}, our open~source implementation of \SYS{} mostly rely on 2 software components---a \textsf{sensor} and a \textsf{power~meter}---which are connected with a \textsc{MongoDB} database.\footnote{\url{https://www.mongodb.com}}
\textsc{MongoDB} offers a flexible and persistent buffer to store input metrics and power estimations.
The \textsf{sensor} is designed as a lightweight process that is intended to run on target nodes with a limited impact.
The \textsf{power~meter} is a remote service that can be deployed whenever needed.
\SYS{} uses this feature to support both online and \emph{post~mortem} power estimations, depending on use cases.

\subsection{Client-side Sensor}\label{sec:sensor}
The component \textsf{sensor} consists in a lightweight software daemon deployed on all the nodes that need to be monitored.

\paragraph*{Static power isolation}
When the node boots, the \textsf{sensor} starts the idle consumption isolation phase (cf. Section~\ref{sec:idle}) by monitoring the PKG and DRAM power consumptions reported by RAPL along the global idle CPU time and the \texttt{fork}, \texttt{exec} and \texttt{exit} process control activities provided by Linux \emph{process information pseudo-filesystem} (procfs).
Whenever a process control activity or the global idle CPU time exceed 99\,\% during this phase, the power samples are discarded to prevent the impact of background activities on the static power isolation process.
As stated in \ref{sec:idle}, this phase is only required when the idle attribution policy consider the idle consumption as a power leakage.
It is not needed to run this phase as long as there is no change in the hardware configuration of the machine (specifically CPU or DRAM changes).

\paragraph*{Event selection}
Once completed, the \textsf{sensor} switches to the event selection phase (cf. Section~\ref{sec:hpc}).
To select the most accurate \emph{Hardware performances counters} to estimate the power of a given node, \SYS{} need to identify the \textsc{HwPC} statistically correlated with the power consumption of the components.
For that, the \textsf{sensor} monitors the power consumption reported by RAPL and the maximum simultaneous \textsc{HwPC} events possible without multiplexing, as it can a significant noise and distort the correlation coefficient of the events, over a (configurable) period of $30$ ticks.
The maximal amount of simultaneous \textsc{HwPC} events depends of the micro-architecture of the CPU and will be detected at runtime using the PMU detection feature of the \emph{libpfm4} library.\footnote{\url{http://perfmon2.sourceforge.net}}
We then correlate the power consumption with the values of the monitored \textsc{HwPC} events and rank them by highest correlation with RAPL and lowest correlation across the other \textsc{HwPC}.
Whenever possible, fixed \textsc{HwPC} event counters are selected in priority to avoid consuming a programmable counter.

\paragraph*{Control groups}\label{sec:cgroups}
\SYS{} leverages the \emph{control groups} (Cgroups) implemented by Linux to support a wide range of monitoring granularities, from single processes, to software containers (\textsc{Docker}),\footnote{\url{https://docker.com}} to virtual machines (using \textsc{Libvirt}).\footnote{\url{https://libvirt.org}}
The \textsf{sensor} also implement a kernel module that is in charge of configuring the Cgroups to monitor the power consumption of kernel and system activities, which is not supported by default.
To do so, this module defines 2 dedicated Cgroups for the roots of the \textsf{system} and the \textsf{kernel} process hierarchy.

\paragraph*{Event monitoring}
Once done with the above preliminary phases, the \textsf{sensor} automatically starts to monitor the selected \textsc{HwPC} events together with RAPL measurements for the DRAM and CPU components at a given frequency and it reports these samples to the \textsc{MongoDB} backend (cf. Section~\ref{sec:monitoring}).
The \textsf{sensor} monitors the selected \textsc{HwPC} events for the host and all the Cgroups synchronously to ensure that all the reported samples are consistent when computing the power models.

\subsection{Server-side Power Meter}\label{sec:meter}
The \textsf{power~meter} is implemented as a software service that requires to be deployed on a single node (\emph{e.g.}, the master of a cluster).
The \textsf{power~meter} can be used online to produce real-time power estimations or offline to conduct \emph{post~mortem} analysis.
This component consumes the input samples stored in the \textsc{MongoDB} database and produces power estimations accordingly.
\SYS{} adopts a modular architecture based on the actor programming model, which we use to integrate a wide range of input/output data storage technologies (MongoDB, InfluxDB, etc.) and to implement power estimations at scale by devoting one actor per power model.

\paragraph*{Power modelling}
The \textsf{power~meter} provides an abstraction to build power models.
In this paper, the power model we report on is handled by \emph{Scikit-Learn}, which is the \emph{de~facto} standard Python library for general-purpose machine learning.\footnote{\url{https://scikit-learn.org}}
We embed the Ridge regression of \emph{Scikit-Learn} in an actor, which is in charge of delivering a power estimation whenever a new sample is fetched from the database.

\paragraph*{Model calibration}
When the error reported by the power model exceeds the threshold defined by the user, the \textsf{power~meter} triggers a new calibration of the power model to take into account the latest samples.
This new power model is checked against the last sample to estimate its accuracy.
If it estimates the power consumption below the configured threshold, then the actor is updated accordingly.

\paragraph*{Power estimation}
Power estimations are delivered at the scale of a node and for the Cgroups of interest.
These scope of these Cgroups can reflect the activity of nodes' kernel and system, as well as any job or service running in the monitored environment.
These power estimations can then be aggregated by owner, service identifier or any other key, depending on use cases.
They can also be aggregated along time to report on the energy footprint of a given software system.

%% file: sections/validation.tex
\section{Validation of \SYS{}}\label{sec:validation}
This section assesses the efficiency and the accuracy of \SYS{} to evaluate the power consumption of running software containers.

\subsection{Evaluation Methodology}\label{sec:testbeds}
We follow the experimental guidelines reported by~\cite{DBLP:journals/corr/abs-1801-02381} to enforce the quality of our results.

\paragraph*{Testbeds \& workloads}
While our production-scale deployments of \SYS{} cover both \textsc{Kubernetes} and \textsc{OpenStack} clusters,
for the purpose of this paper, we chose to report on more standard benchmarks, like \textsc{Stress\,NG}\footnote{\url{https://launchpad.net/stress-ng}} and NASA's \emph{NAS Parallel Benchmarks} (NPB)~\cite{Bailey:1991} to highlight the benefits of our approach.
Our setups are reproduced on the \textsc{Grid5000} testbed infrastructure,\footnote{\url{https://www.grid5000.fr}} which provides multiple clusters composed of powerful nodes.
In this evaluation, we use a Dell PowerEdge C6420 server having two Intel® Xeon® Gold 6130 Processors (Skylake) and 192\,GB of memory (12 slots of 16\,GB DDR4 2666MT/s RDIMMs).
We are using the Ubuntu 18.04.3 LTS Linux distribution running with the 4.15.0-55-generic Kernel version, where only a minimal set of daemons are running in background.
As stated in \ref{sec:cgroups}, we are using the Cgroups to monitor the activity of the running processes independently.
In the case of the system services managed by systemd and the services running in Docker containers, their Cgroups membership is automatically handled as part of their lifetime management.

For this host, the reported TDP for the CPU is 125\,Watts and 26\,Watts for the DRAM.
Theses values were obtained from the \texttt{PKG\_POWER\_INFO} and \texttt{DRAM\_POWER\_INFO} \emph{Model Specific Registers} (MSR).
The energy and performance optimization features of the CPU---\emph{i.e.}, \emph{Hardware P-States} (HWP), \emph{Hyper-Threading} (HT), \emph{Turbo Boost} (TB) and \emph{C-states}, are fully enabled and use the default configuration of the distribution.
The default CPU scaling driver and governor for the distribution are \emph{intel\_pstate} and \emph{powersave}.

In all our experiments, we configure \SYS{} to report power measurements twice a second ($2\,Hz$) with an error threshold of 5\,Watts for the PKG and 1\,Watt for the DRAM.

\paragraph*{Objectives}
We evaluate \SYS{} with the following criteria:
\begin{itemize}
	\item The \emph{quality} of the power estimations when running sequential and parallel workloads;
	\item The \emph{accuracy and stability} of the power models across different workloads;
	\item The \emph{overhead} of the \SYS{} \textsf{sensor} component on the monitored host.
\end{itemize}

\paragraph*{Reproducibility}
For the sake of reproducible research, \SYS{}, the necessary tools, deployment scripts and resulting datasets are open-source and publicly available on GitHub.\footnote{\SYSURL{}}

\subsection{Experimental Results}\label{sec:metrics}
\paragraph*{Quality of estimations}

\begin{figure*}
	\centering
	\includegraphics[width=\linewidth]{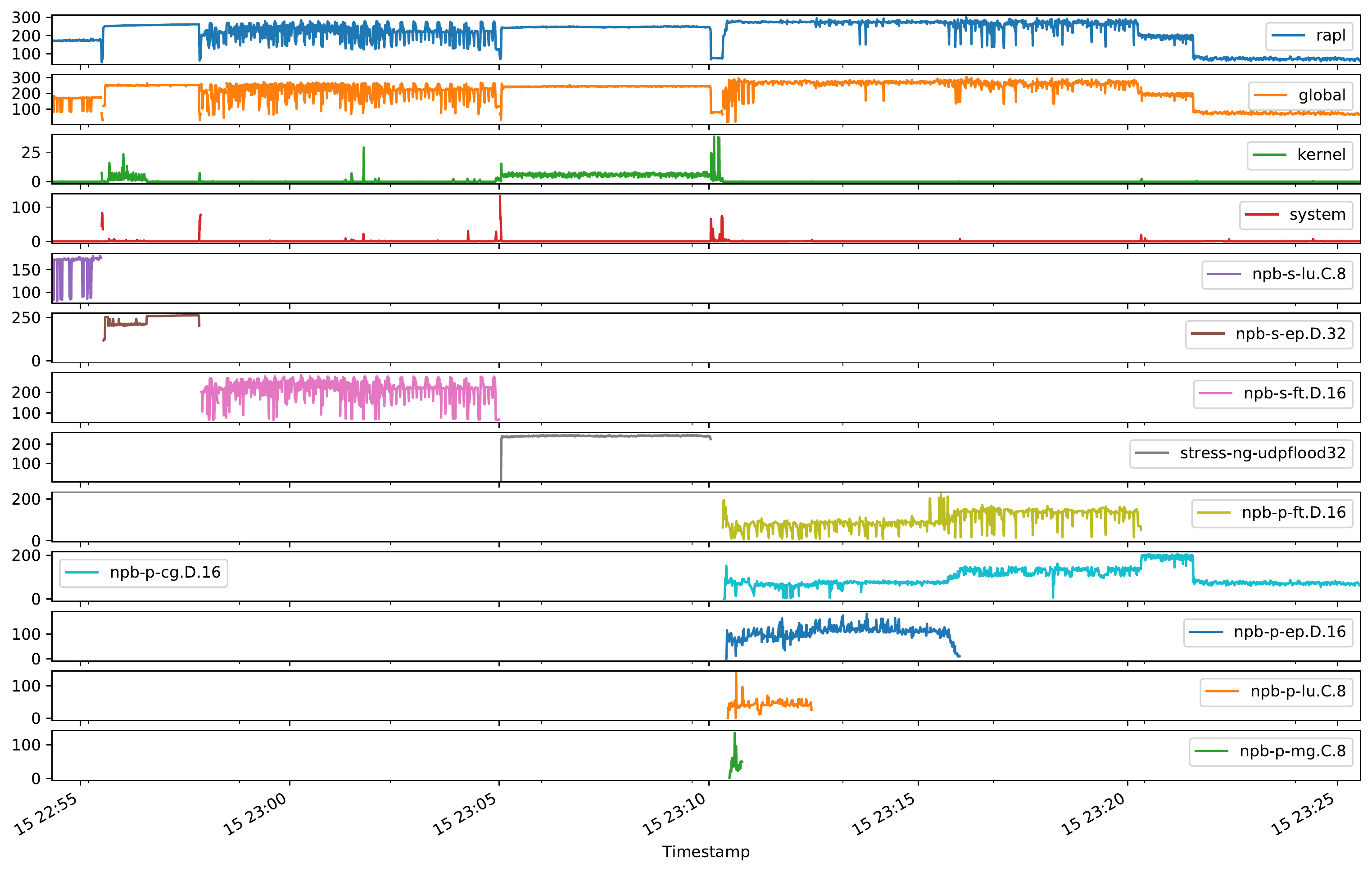}
	\caption{Evolution of the PKG \& DRAM power consumption along time and containers}
	\label{fig:agg-power}
\end{figure*}

\begin{figure}
	\centering
	\includegraphics[width=\linewidth]{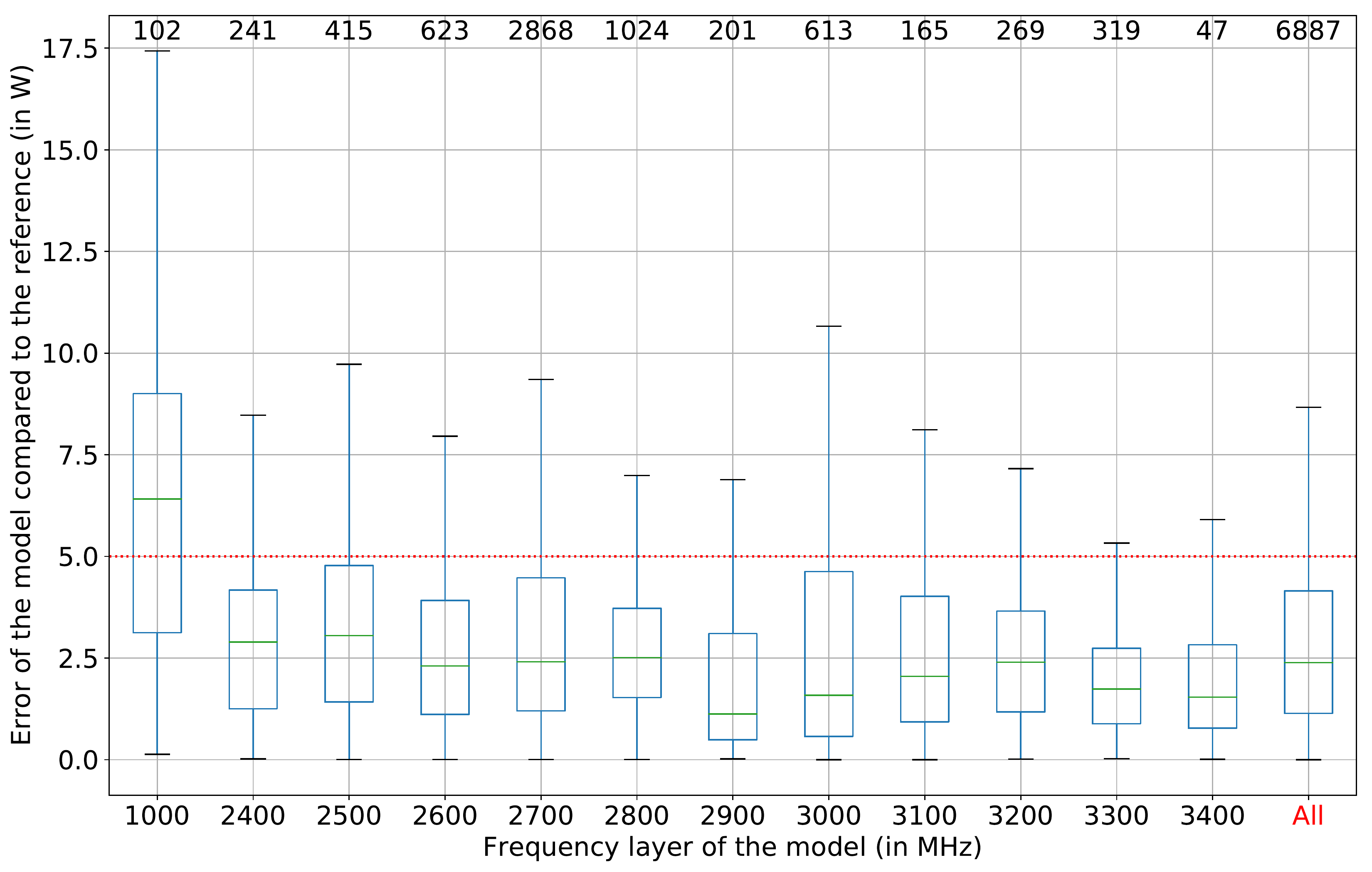}
	\caption{Global \& per-frequency error rate of the PKG power models}
	\label{fig:cpu-error}
\end{figure}

\begin{figure}
	\centering
	\includegraphics[width=\linewidth]{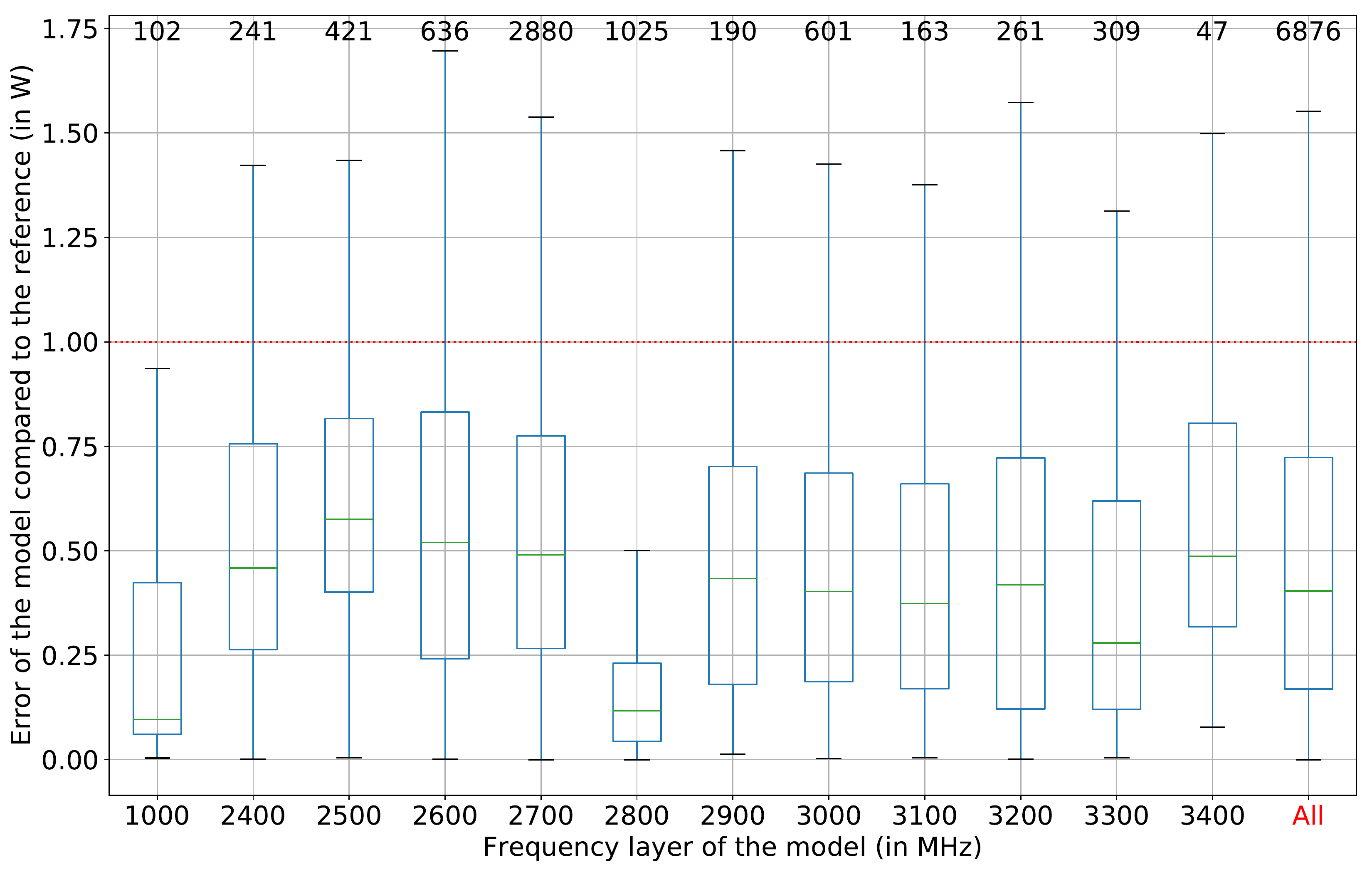}
	\caption{Global \& per-frequency error rate of the DRAM power models}
	\label{fig:dram-error}
\end{figure}

Figure~\ref{fig:agg-power} first reports on the PKG and DRAM power consumptions we obtained with \SYS{}.
The first line (\textsf{rapl}) refers to the ground truth power measurements we sample for the PKG and the DRAM via the \textsc{HwPC} events \texttt{RAPL\_ENERGY\_PKG} and \texttt{RAPL\_ENERGY\_DRAM}, respectively.
The second line (\textsf{global}) refers to the power measurements estimated by \SYS{} for the PKG and the DRAM components from \texttt{CPU\_CLK\_THREAD\_UNHALTED:REF\_P}, \texttt{CPU\_CLK\_THREAD\_UNHALTED:THREAD\_P}, \texttt{INSTRUCTIONS\-\_RETIRED} (fixed counters), and \texttt{LLC\_MISSES} (programmable counter).
The list of events has been automatically selected by the \textsf{sensor} component as presenting the best correlation with RAPL samples, as described in Section~\ref{sec:hpc}.
The error for each of the power models are further discussed in Figures~\ref{fig:cpu-error} and~\ref{fig:dram-error}.

The lines \textsf{kernel} and \textsf{system} isolates the power consumption induced by all kernel and system activities.
Kernel activities include devices specific background operations, such as \emph{Network interface controller} (NIC) and disks I/O processing queues, while system activities covers the different services, like the SSH server and Docker daemon, running on the node.

The remaining lines reports on individual power consumptions of a set of NPB benchmarks, which are executed in sequence (\texttt{lu}, \texttt{ep}, \texttt{ft}) or concurrently (\texttt{ft}, \texttt{cg}, \texttt{ep}, \texttt{lu}, \texttt{mg}) with variable number of cores (ranging from 8 to 32 cores).
One can observe that \SYS{} supports the isolation of power consumptions at process-level by leveraging Linux Cgroups.
This granularity allows \SYS{} to monitor indifferently processes, containers or virtual machines.

\begin{figure}
	\centering
	\includegraphics[width=\linewidth]{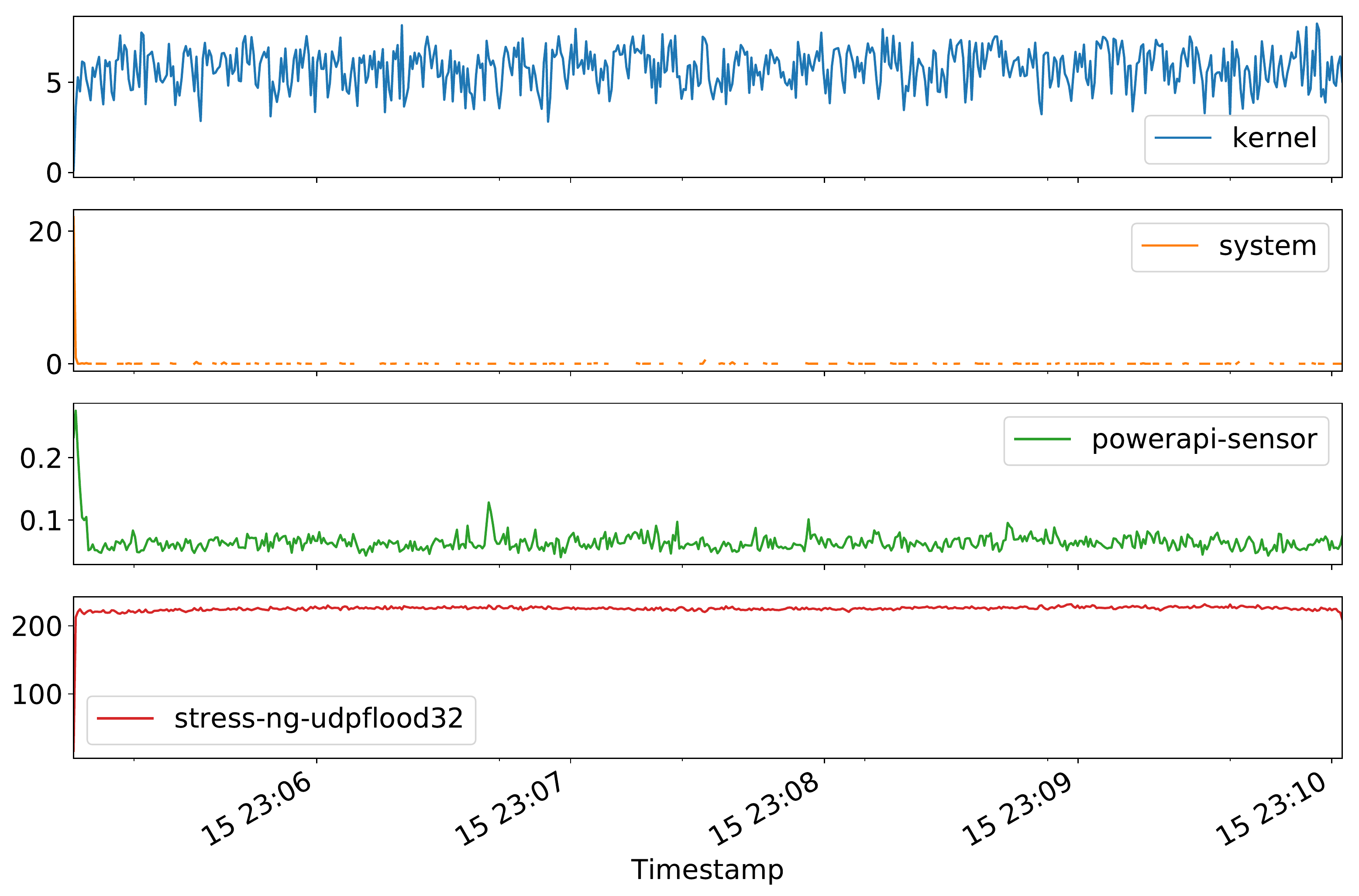}
	\caption{Illustrating the activity of the kernel when flooding UDP}
	\label{fig:stress-ng}
\end{figure}

We also run \texttt{stress-ng} to observe potential side effects on the kernel activity by starting 32 workers that attempt to flood the host with UDP packets to random ports (cf. Figure~\ref{fig:stress-ng}).
While it remains negligible compared to the power consumption of the UDP flood process (2.971\,W vs. 120.322\,W on average), one can observe that this stress induces a lot of activity at the kernel to handle IO, while the rest system is not severely impacted.

One can also observe that our sensor induces a negligible overhead (less than $0.2$ Watts) with regards to the consumption of surrounding activities.

\paragraph*{Estimation accuracy}
Figures~\ref{fig:cpu-error} and~\ref{fig:dram-error} reports on the distribution of estimation errors we observed per frequency and globally (right part of the plots) for the above scenario.
We also report on the number of estimations produced for each of the frequency (upper part of the plots).
While the error threshold for CPU and DRAM is set to 5\,Watts and 1\,Watts, one can observe that \SYS{} succeeds to estimate the power consumption with less than 4\,Watts and 0.5\,Watt of error for the PKG and DRAM components, respectively.
The only case where estimation error grows beyond this threshold refers to the frequency \textsf{1000\,Hz} of the CPU (cf. Figures~\ref{fig:cpu-error}).%
The frequency \textsf{1000\,Hz} refers to the idle frequency of the node and the sporadic triggering of activities in this frequency induces a chaotic workload which is more difficult to capture for \SYS{} given the limited number of samples acquired in this frequency (102 samples against 2868 samples for the frequency \textsf{2700\,Hz}).

The DRAM component, however, provides a more straightforward behavior to model with the selected \textsc{HwPC} events and therefore reports an excellent accuracy, no matter the operating frequency of the CPU package (cf. Figure~\ref{fig:dram-error}).

The accuracy of the power models generated by \SYS{} are further detailed in Table~\ref{table:models-estimations-accuracy}.
While our approach succeeds to deliver accurate estimations of the power consumption for both CPU and DRAM components, the maximum error refers to the bootstrapping phase of the sensor that requires to acquire a sufficiently representative number of samples in order to build a stable and accurate power model.

\begin{table}[h!]
	\centering
	\caption{Per-socket PKG \& DRAM power models accuracy}
	\label{table:models-estimations-accuracy}
	\resizebox{\columnwidth}{!}{%
	\begin{tabular}{ll|rrrr}
		\toprule
		Resource & Socket & $\varepsilon_{min}$ & $\varepsilon_{max}$ & $\varepsilon_{mean}$ & $\varepsilon_{std}$ \\
		\midrule
		\midrule
		PKG  & 0 &  0.000\,W &  123.888\,W &  3.337\,W &  5.071\,W \\
			 & 1 &  0.002\,W &  103.893\,W &  3.278\,W &  4.459\,W \\
		\midrule
		DRAM & 0 &  0.000\,W &   89.600\,W &  0.577\,W &  2.403\,W \\
  			 & 1 &  0.000\,W &   39.702\,W &  0.600\,W &  1.270\,W \\
		\bottomrule
	\end{tabular}
	}%
\end{table}

\paragraph*{Model stability}
Beyond the capability to accurately estimate the power consumption of software containers, we are also interested in assessing the capability of \SYS{} to generate stable power models over time.
Tables~\ref{table:cpu-models-stability} and~\ref{table:dram-models-stability} therefore reports, for each frequency, on metrics about the stability of power models.
In particular, we look at the number of correct estimations produced by the power models in a given frequency.
Given our input workloads, we can observe that \SYS{} succeeds to reuse a given power model up to $592$ estimations, depending on frequencies.
While we observed that the stability of our power models strongly depends on the sampling frequency, the error threshold, as well as the input workloads, one should note that the overhead for calibrating a power model in a given frequency does not take more than a couple milliseconds, which is perfectly acceptable when monitoring software systems in production.

\begin{table}[h!]
	\centering
	\caption{PKG power models stability per frequency}
	\label{table:cpu-models-stability}
	\resizebox{\columnwidth}{!}{%
	\begin{tabular}{l|r|r|rrrr}
		\toprule
		Frequency & models & total & min & max & mean & std \\
		\midrule
		\midrule
		1000            &      86 &    102 &    1 &   24 &   1.545 &   2.899 \\
		2400            &      38 &    241 &    1 &   30 &   5.738 &   7.404 \\
		2500            &      63 &    415 &    1 &   50 &   4.414 &   6.778 \\
		2600            &      59 &    623 &    1 &   62 &   5.417 &   9.881 \\
		2700            &     392 &   2868 &    1 &  592 &   4.961 &  26.175 \\
		2800            &      47 &   1024 &    1 &  271 &  14.840 &  44.907 \\
		2900            &      21 &    201 &    1 &  107 &   7.730 &  20.535 \\
		3000            &     132 &    613 &    1 &  171 &   4.347 &  15.030 \\
		3100            &      27 &    165 &    1 &   72 &   6.600 &  13.898 \\
		3200            &      43 &    269 &    1 &  126 &   6.255 &  19.509 \\
		3300            &      35 &    319 &    1 &   90 &   8.861 &  19.585 \\
		3400            &       8 &     47 &    1 &   22 &   5.875 &   7.180 \\
		\bottomrule
	\end{tabular}
	}%
\end{table}

\begin{table}[h!]
	\centering
	\caption{DRAM power models stability per frequency}
	\label{table:dram-models-stability}
	\resizebox{\columnwidth}{!}{%
	\begin{tabular}{l|r|r|rrrr}
		\toprule
		Frequency & models & total & min & max & mean & std \\
		\midrule
		\midrule
		1000            &      27 &    102 &    2 &   38 &  11.333 &  12.103 \\
		2400            &      17 &    241 &    1 &   44 &  11.476 &  11.470 \\
		2500            &      34 &    421 &    1 &   87 &   6.682 &  12.068 \\
		2600            &      67 &    636 &    1 &   95 &   6.913 &  13.754 \\
		2700            &     280 &   2880 &    1 &  538 &   9.260 &  37.811 \\
		2800            &      19 &   1025 &    1 &  349 &  29.285 &  83.863 \\
		2900            &      21 &    190 &    1 &   35 &   7.037 &   9.146 \\
		3000            &      46 &    601 &    1 &   85 &  10.732 &  15.487 \\
		3100            &      20 &    163 &    1 &   48 &   8.150 &  12.533 \\
		3200            &      27 &    261 &    1 &   42 &   9.666 &  11.187 \\
		3300            &      27 &    309 &    1 &   78 &  11.444 &  16.158 \\
		3400            &      11 &     47 &    1 &   10 &   4.272 &   3.635 \\
		\bottomrule
	\end{tabular}
	}%
\end{table}

\paragraph*{Monitoring overhead}
Regarding the runtime overhead of \SYS{}, one can observe in Figure~\ref{fig:stress-ng} that the power consumption of \SYS{} is negligible compared to the hosted software containers.
To estimate this overhead, we leverage the fact that the \textsf{sensor} component is running inside a software container, thus enabling \SYS{} to estimate its own power consumption.
In particular, one can note in Table~\ref{table:sensor-power-consumption} that the \textsf{sensor} power consumption represents 1.2\,Watts for the PKG and 0.06\,Watts for the DRAM, on average, when running at a frequency of $2\,Hz$.
The usage of the \emph{Hardware Performance Counters} (HwPC) is well known for its very low impact on the observed system, hence it does not induce runtime performance penalties~\cite{Colmant:2015,Dolz:2015,Kurpicz:2016,Prekas:2015}.
Additionally, we carefully took care of the cost of sampling these HwPC events and executing as little as possible instructions on the monitored nodes.

\begin{table}[h!]
	\centering
	\caption{Per-component power consumption of the sensor}
	\label{table:sensor-power-consumption}
	\resizebox{.8\columnwidth}{!}{%
	\begin{tabular}{l|rrrr}
		\toprule
		Power     & min &        max &      mean &       std \\
		\midrule
		\midrule
		PKG   &  0.0\,W &  52.078\,W &  1.241\,W &  6.559\,W \\
		DRAM  &  0.0\,W &  29.966\,W &  0.065\,W &  0.566\,W \\
		\bottomrule
		\end{tabular}
	}%
\end{table}

By proposing a lightweight and packaged software solution that can be easily deployed across monitored hosts, we facilitate the integration of power monitoring in large-scale computing infrastructures.
Futhermore, the modular architecture of \SYS{} can accommodate existing monitoring infrastructures, like \textsc{Kubernetes Metrics} or \textsc{OpenStack Ceilometer}, to report on the power consumption of applications.
The following section therefore demonstrates this capability by deploying a distributed case study atop of a \textsc{Kubernetes} cluster.

\subsection{Tracking the Energy Consumption of Distributed Systems}
\begin{figure}
	\centering
	\includegraphics[width=\linewidth]{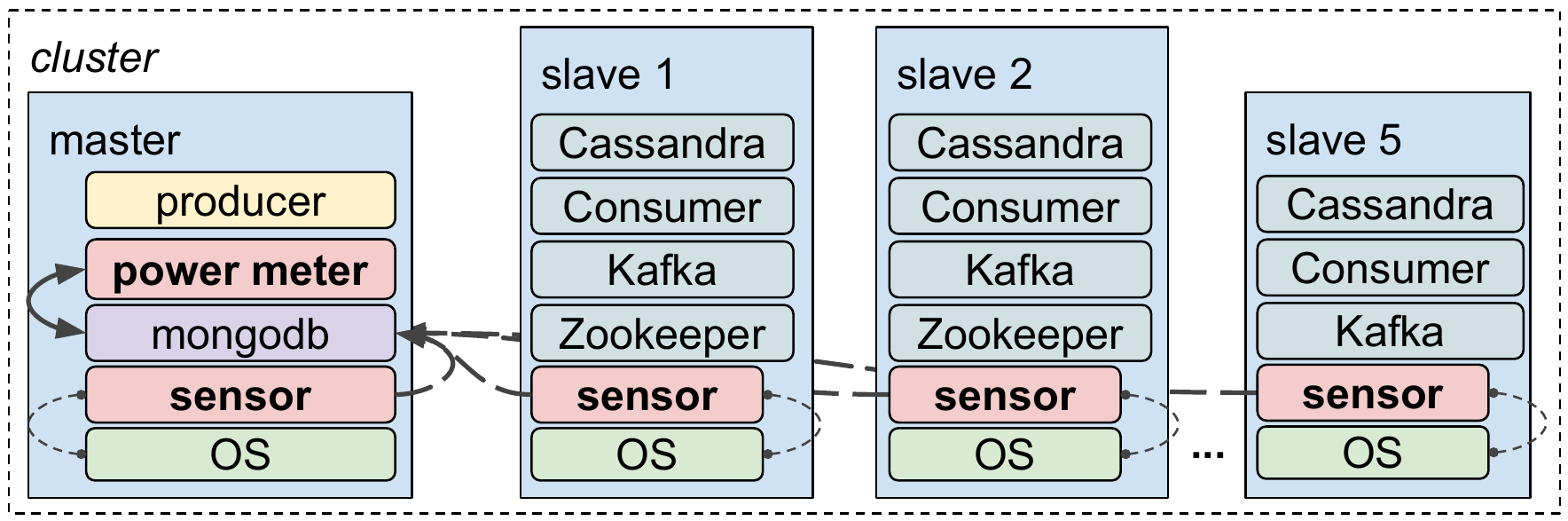}
	\caption{Deployment of Kubernetes IoT backend services across 6 nodes}
	\label{fig:smartwatts-kafka}
\end{figure}

To further illustrate the capabilities of \SYS{}, we take inspiration from~\cite{DBLP:conf/ccgrid/ColmantFRS17} to deploy a distributed software systems that processes messages forwarded by IoT devices to a pipeline of processors connected by a \textsc{Kafka} cluster to a \textsc{Cassandra} storage backend.
Figure~\ref{fig:smartwatts-kafka} depicts the deployment of this distributed system on a \textsc{Kubernetes} cluster composed of 1 master and 5 slave nodes.
The input workload consists in a producer injecting messages in the cluster with a throughput ranging from 10 to 100~MB/s.

\begin{figure}
	\centering
	\includegraphics[width=\linewidth]{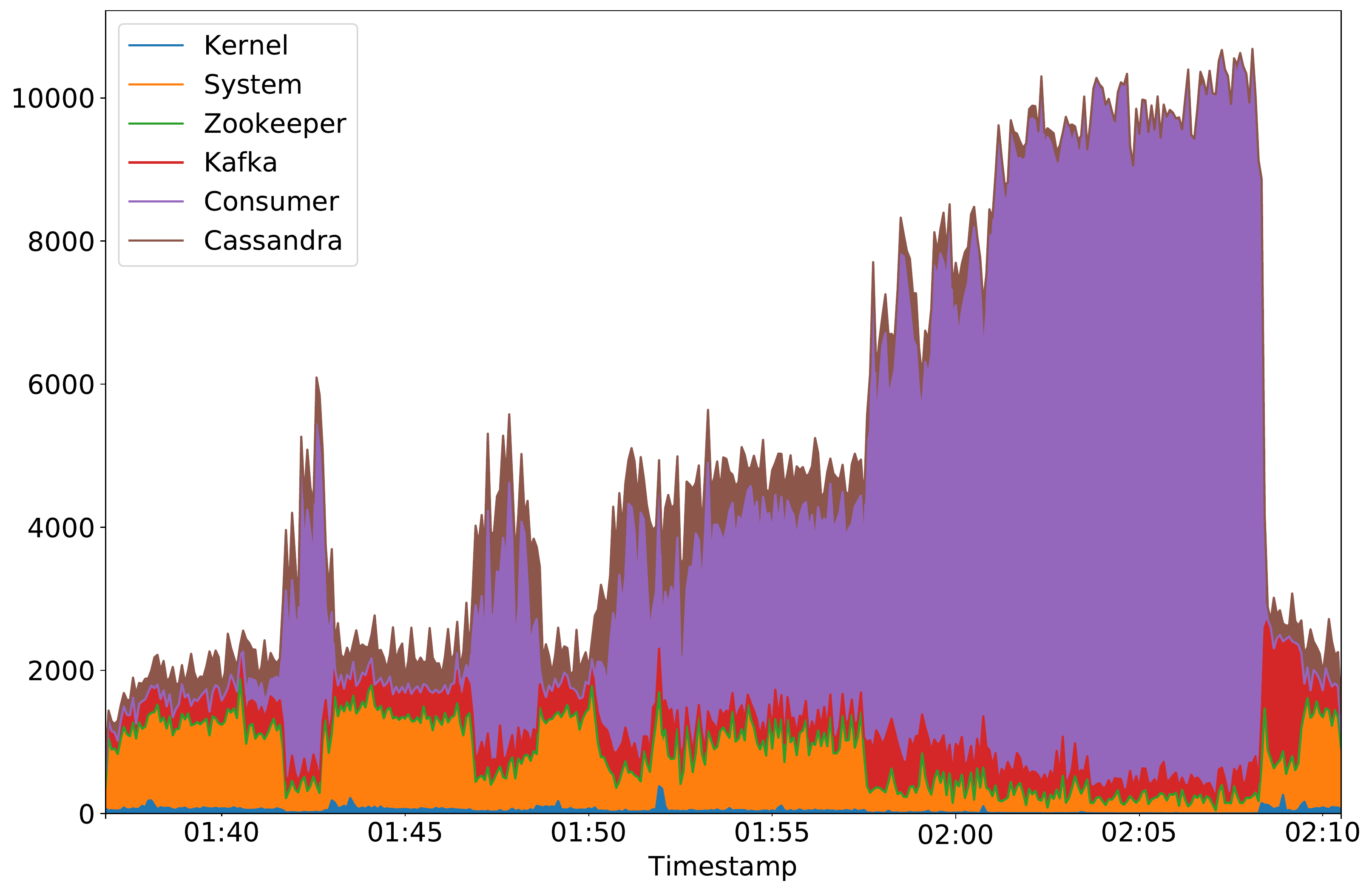}
	\caption{Monitoring of service-level power consumptions}
	\label{fig:distributed-pipeline-targets}
\end{figure}

Figure~\ref{fig:distributed-pipeline-targets} reports on the evolution of the power consumption per service while injecting the workload from the master node.
One can observe that, when increasing the message throughput, the most impacted service is the \textsf{Consumer}, which requires extensive energy to process all the messages enqueued by the \textsf{Kafka} service.
This saturation of the \textsf{Consumer} service seems to represent a core bottleneck in the application.

\begin{figure*}
	\centering
	\includegraphics[width=\linewidth]{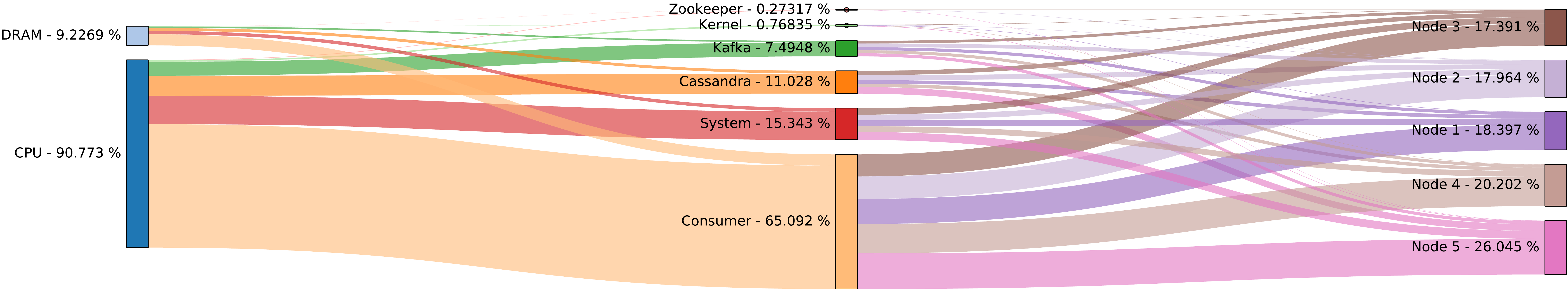}
	\caption{Distribution of the energy consumption across nodes and resources}
	\label{fig:distributed-pipeline-energy}
\end{figure*}

To further dive into this problem, we consider another perspective on the deployment in order to investigate the source of this efficiency limitation.
While the execution of this workload requires 1.32 MJoules of energy to process the whole dataset, Figure~\ref{fig:distributed-pipeline-energy} further dives inside the distribution of the energy consumption of individual pods along the PKG and DRAM components as a Sankey diagram~\cite{LUPTON2017141}.
This diagram builds on the capability of \SYS{} to aggregate power estimations along time to report on the energy consumption, as well as its capacity to track power consumption from software processes (on left-hand side) down to hardware components (on the right-hand side).
This diagram can therefore be used to better understand how a distributed software system takes advantage of the underlying hardware components to execute a given workload.
In particular, one can observe that 91\,\% of the energy is spent by the CPU package, while the \textsf{Consumer} service drains 65~\% of the energy consumption of the monitored scenario.
Interestingly, one can observe that this energy consumption is evenly distributed across the 5 slaves, thus fully benefiting from the pod replication support of \textsc{Kubernetes}.
The observed energy overhead is not due to the saturation of a single node, but rather seems to be distributed across the nodes, therefore highlighting an issue in the code of the \textsf{Consumer} service.
This issue is related to the acknowledgement of write requests by the \textsc{Cassandra} service, which prevents the \textsc{Consumer} service to process pending messages.

We believe that, thanks to \SYS{}, system administrators and developers can collaborate on identifying energy hotspots in their deployment and adjusting the configuration accordingly.

%% file: sections/conclusion.tex
\section{Conclusion}\label{sec:conclusion}
Power consumption is critical concern in modern computing infrastructures, from clusters to data~centers.
While the state of practice offers tools to monitor the power consumption at a coarse granularity (\emph{e.g.}, nodes, sockets), the literature fails to propose generic power models, which can be used to estimate the power consumption of software artefacts.

In this paper, we therefore reported on a novel approach, named \SYS{}, to deliver per-container power estimations for PKG and DRAM components.
In particular, we propose to support self-calibrating power models to estimate the  PKG and DRAM power consumption of software containers.
Unlike static power models that are trained for a specific workload, our power models leverage sequential learning principles to be adjusted online in order to match unexpected workload evolutions and thus maximize the accuracy of power estimations.

While we demonstrate this approach using Intel RAPL and the Linux's \emph{perf\_events} interface, we strongly believe that it can be used as a solid basis and generalized to other architectures and system components.
In particular, we are working on the validation of our approach with AMD Ryzen architecture (including a support for RAPL).

Thanks to \SYS{}, system administrators and developers can monitor the power consumption of individual containers and identify potential optimizations to apply in the distributed system they manage.
Instead of addressing performance issues by adding more resources, we believe that \SYS{} can favorably contribute to increase the energy efficiency of distributed software systems at large.